\documentclass[aps,showpacs,pre,twocolumn,amsfonts,floats]{revtex4}
\usepackage{graphicx}
\begin{document}

\title{Mode-Locking of mobile discrete breathers}
\author{D.~Zueco$ ^{(a),(c),(d)}$,P.~J.~Mart\'{\i}nez$^{(b),(c),(d)}$
L.M. Flor\'{\i}a$^{(a),(c),(d)}$, and F.~Falo$^{(a),(c),(d)}$}
\affiliation{$^{(a)}$ Dept.\ de F\'{\i}sica de la Materia Condensada,
  Universidad de Zaragoza, 50009~Zaragoza, Spain}
\affiliation{$^{(b)}$ Dept.\ de F\'{\i}sica Aplicada, Universidad de
  Zaragoza, 50009~Zaragoza, Spain}
\affiliation{$^{(c)}$ Dpto.\ de Teor\'{\i}a y Simulaci\'on de Sistemas
  Complejos.
  Instituto de Ciencia de Materiales de Arag\'on ICMA,
  C.S.I.C. - Universidad de Zaragoza, 50009~Zaragoza, Spain}
\affiliation{$^{(d)}$ Instituto de Biocomputaci\'on y F\'{\i}sica de Sistemas
  Complejos BIFI, Universidad de Zaragoza, 50009~Zaragoza, Spain}
\date{\today}
\begin{abstract}

We study numerically synchronization phenomena of mobile discrete
breathers in dissipative nonlinear lattices periodically forced. When
varying the driving intensity, the breather velocity generically locks
at rational multiples of the driving frequency. In most cases, the
locking plateau coincides with the linear stability domain of the
resonant mobile breather and the desynchronization occurs by regular
appearance of type I intermittencies. However, some plateaux also show
chaotic mobile breathers with locked velocity in the locking
region. The addition of a small subharmonic driving tames the locked
chaotic solution and enhances the stability of resonant mobile
breathers.
  
\end{abstract}
\maketitle


\section {INTRODUCTION}

Nonlinear lattices provide some of the most interesting model systems
of macroscopic nonlinear behavior with experimental realizations \cite
{Scott}. From a theoretical perspective they have been progressively
recognized not as mere discretizations of nonlinear continuous
fields (unavoidable for numerical computations), but as a target of
interest by themselves, due to the distinctive features associated to
{\em discreteness}.

\begin{figure}
  \begin{center}(a)
    \includegraphics[angle = -90, width=0.4\textwidth]{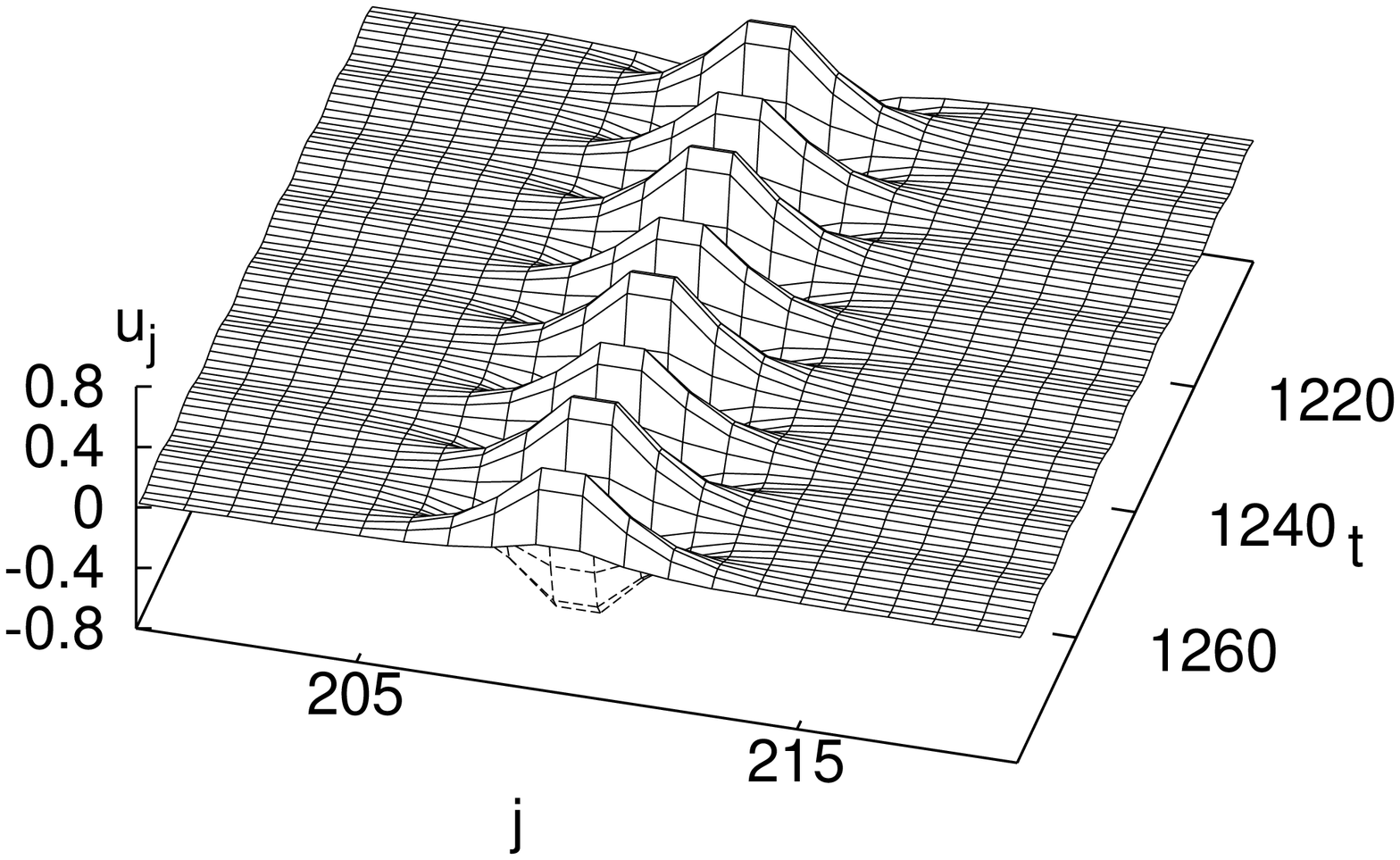}
    \\
    (b)
    \includegraphics[angle = -90, width=0.4\textwidth]{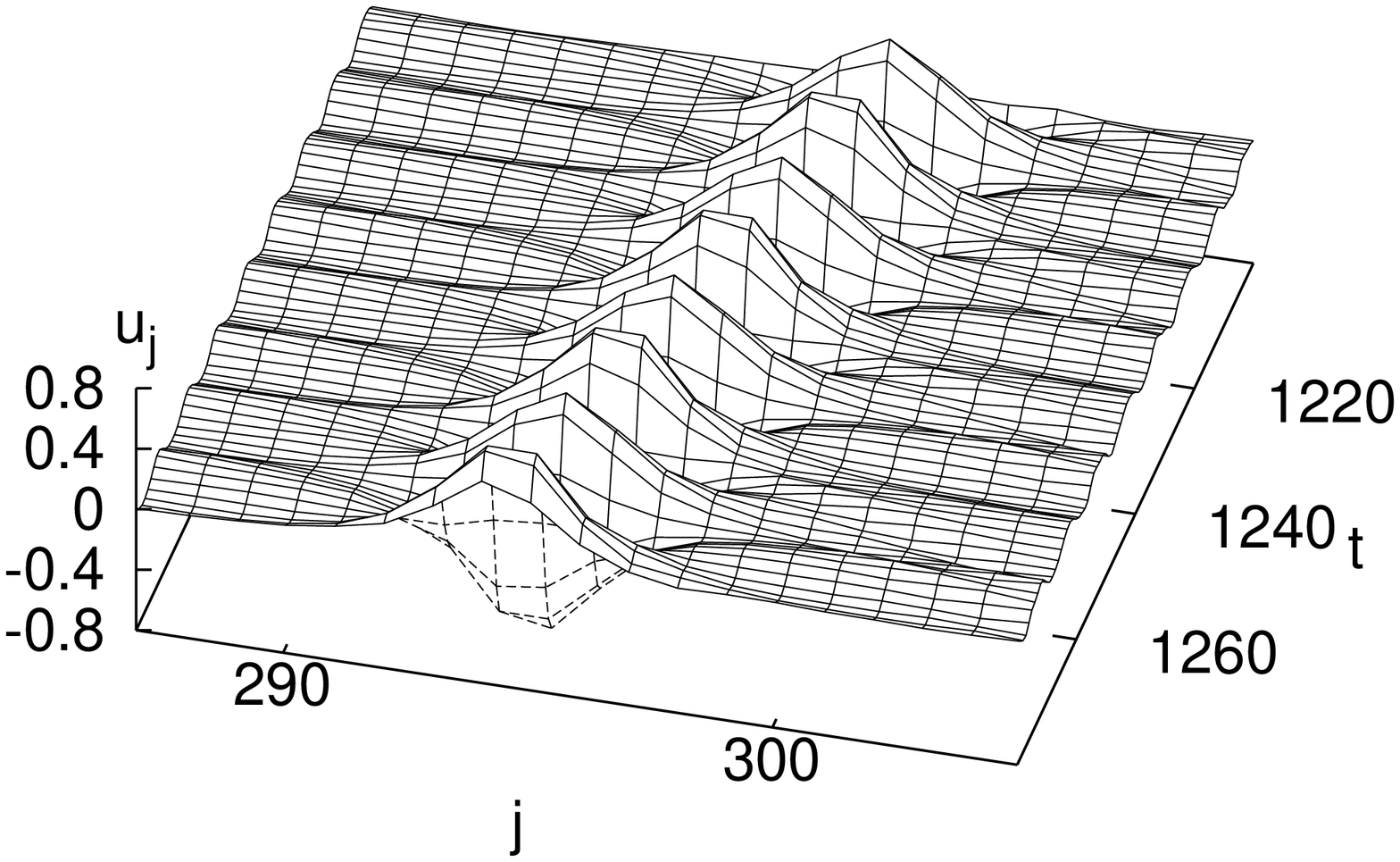}
    \caption{Time evolution of two discrete breathers: (a) periodic
    pinned  breather, (b) $1/2$-resonant (see text) mobile breather.
    The localization center moves a lattice site every two periods of
    the external force}
    \label{fig3D}
  \end{center}
\end{figure}

Among the variety of behaviors of the lattice nonlinear dynamics, we
focus our attention here on the called ``intrinsic localized modes''
or ``discrete breathers'' (DB's for short). DB's are exact periodic,
large amplitude and exponentially localized solutions
\cite{FW98}. These solutions are made possible by the combination of
nonlinearity and discreteness: Nonlinearity allows for solutions out
of the linear mode bands, due to the frequency dependence of the
oscillation amplitude. On the other hand, discreteness sets an upper
cutoff in the band structure and prevents the multi-harmonic
resonances of DB with extended linear modes. These two simple
ingredients are enough for the existence of DB's, wherefrom the
generality and wide range of interest of the phenomenon. To visualize
an immobile DB see the upper fig. \ref{fig3D}.

These excitations are not only interesting from a theoretical point of
view but with respect to the experimental applicability as well,
concerning fields as diverse as biophysics (myelinated nerve fibers
\cite{Scott}, biopolymer chains \cite{ibanes}), nonlinear optics
(photonic crystals and waveguides \cite{mcgurn}), Josephson effect
(superconducting devices \cite{trias, binder}, Bose-Einstein
condensates \cite{cataliotti}) or the physics of glassiness
\cite{elefth2003}. This makes discrete breathers to be the object of a
remarkable multidisciplinary interest.  

Unlike localization due to impurities or disorder (Anderson), 
intrinsic localization phenomena support mobile DB solutions, {\it
  i.e.}, exponentially localized oscillations where the localization
center propagates along the lattice as time goes by 
(see the lower fig. \ref{fig3D}). Although rigorous results apply to 
the generic existence of non-moving DB, much less is known about the 
conditions for their mobility. In this respect, Hamiltonian 
discrete breathers have received (comparatively) more attention 
than their dissipative counterparts. However, from the 
perspective of applications to experimental situations, the 
unavoidable coupling (both thermal and non-thermal) of the relevant 
degrees of freedom to a variety of other ones, often demands to 
consider open systems where power balance, instead of energy 
conservation, governs the nonlinear dynamics of the lattice.

In this article we pay attention to the problem of synchronization and
resonant behavior in {\em mobile} discrete breathers (MB's for short) of
the forced and damped sine-Gordon lattice (the 1d standard
Frenkel-Kontorova model \cite{martinez03encyclopedia,braun,floria96}),
illustrating the effects of {\em time scales competition} in breather
dynamics, in a discrete, dissipative and non-integrable context. The
two time scales of the moving pulse are respectively associated to its
frequency, $\omega_b$, and its velocity $v_b$. Our main results can be
summarized as follows:

\begin{itemize}
\item[{\em (i)}] Locking of the breather mean velocity at some 
  rational values of the ratio ${2\pi v_b}/{\omega_b}$ 
  for ranges of parameters (coupling, driving strength, 
  {\ldots} ) is observable. 
  
\item[{\em (ii)}] The synchronization of breather velocity is 
  deeply rooted in the structural stability of pure resonant 
  (to be defined soon) MB's, but it is by no means limited to 
  it: The locking island in parameter space is generally larger 
  than the linear stability domain of the pure resonant breather 
  state. 
  
\item[{(iii)}] The ``extra'' locking domain is characterized by 
  more complex attracting breathers, sometimes chaotic in the 
  breather core, but still keeping a locked velocity at large 
  integration times.

\item[{(iv)}] When (sub)-harmonic perturbations are added to the 
  driving term, the stability of the pure resonant MB is enhanced 
  (sometimes substantially), taming chaotic dynamics and enlarging 
  the locking step size.

\end{itemize}

The paper is organized as follows: After this introductory section 
we present in section II, in a brief but self-contained manner, the 
relevant and most basic aspects of dissipative breather mobility in 
the forced and damped discrete sine-Gordon equation (Frenkel-Kontorova 
model). The definition and characterization of 
($p/q$)-resonant mobile DB's along with the extended Floquet method 
for the analysis of their linear stability are explained in this section.

In section III we present our numerical results. We show that 
steps of mode-locking velocity, where ($p/q$)-resonant solutions 
exist, are found when varying the driving strength. We see 
that this phenomenon is quite general since it is also found 
for an open set of coupling parameter values. In section IV we 
focus attention on the unlocking transition, {\em i.e.} the 
transition from ($p/q$)-resonant locking state to quasiperiodic 
(irrational ${2\pi v_b}/{\omega_b}$) generic velocity. This 
transition is characterized as a bifurcation via intermittencies 
of type I by using the Floquet methods reviewed in section II. 
Section V is devoted to the phenomenon of locking enhancement by 
adding small additional subharmonic driving. 
Finally, some concluding remarks are given in section VI.

\section{Mobile DB in the F-K model. Resonant states and their stability.}
\label{sec:model}

Mobile dissipative discrete breathers have been well-characterized 
\cite{martinez03} in the standard Frenkel-Kontorova chain subject 
to homogeneous periodic driving and viscous damping (see below). 
These solutions are {\em attractors} of the dynamics, and thus 
they are surrounded in phase space by a basin of attraction of 
initial conditions. This fact not only provides fast and accurate 
numerical methods for continuation of generic mobile breathers 
(in contrast with expensive root finding methods for continuation 
in the Hamiltonian case), but also guarantees the very existence 
\cite{mackay98} of exact moving breathers (in contrast also with 
the Hamiltonian situation, where the stability and generality of 
exact moving solutions is currently an issue of debate \cite{chaos}). 

Generic MB's unavoidably excite extended modes (loosely referred to 
as phonons) which tend to delocalize energy. However, in dissipative 
systems, the locally excited phonons decay exponentially so that the 
mobile breather keeps on a finite localization length, essentially 
determined by the self-generated phonon dressing. The power spectrum 
(Fourier) analysis of the numerically exact MB's nicely validates 
their description as a moving source of damped radiation to the 
extent that the predictions of the theory exactly match the numerical 
spectra \cite{MFMF01}. For resonant states (to be defined below) one 
can use Floquet methods in order to perform a more thorough analysis 
of these examples of exact non-integrable mobility. This will become 
technically precise along this section, after explaining briefly the 
model and some relevant issues for the sake of self-containedness. 

The equations of motion of the standard Frenkel-Kontorova model 
subject to a harmonic driving force 
$F_{ac}\sin(\omega_b t)$ and a viscous damping $\alpha$, are, 
in dimensionless form:
\begin{eqnarray}
  \nonumber\ddot{u}_j&+&\alpha\dot{u}_j+\frac{1}{2\pi}\sin(2\pi
u_j)
\\
&&=
C\left(u_{j+1}-2u_j+u_{j-1}\right) + F_{ac}\sin(\omega_b t)
  \label{eq:mot}
\end{eqnarray}
where $C$ denotes the coupling (indeed the coupling/nonlinearity 
ratio) between neighboring nonlinear oscillators $u_j(t)$ of unit 
mass. 

Two different mechanisms for mobility of DB solutions of equation 
(\ref{eq:mot}) have been observed. 
\begin{itemize}
\item[ i)] The spontaneous mirror symmetry-breaking of 
  pinned discrete breathers (DB), that occurs at moderately 
  low couplings, pave the way to mobility in a very 
  natural manner, because a moving DB is a solution with 
  broken symmetry. Indeed, this simple idea is at the 
  origin of a very useful procedure 
  \cite{MFMF01,CAT96,AC98} to prepare good 
  initial conditions in the basin of attraction of exact 
  mobile DB's: As described in detail elsewhere \cite{martinez03}, 
  adding a small perturbation (along the symmetry-breaking 
  eigenvector, often dubbed as depinning mode) to the 
  immobile exact DB often evolves asymptotically to an 
  attracting moving solution. 

\item[ ii)] Immobile quasiperiodic DB's have 
  been seen to suffer from depinning parametric instabilities 
  leading to mobile (typically slower) DB's \cite{martinez03} in 
  a range of somewhat larger values of the coupling parameter. 
  No stable pinned DB coexists in this range with moving solutions.
\end{itemize}

Once an attracting MB has been precisely determined, 
continuation from it through variations of parameters such as 
coupling or driving intensity, generates a numerically 
continued branch of moving breathers. The continuation 
procedure of MB's from an initial one proceeds by a 
very small (adiabatic) change of a parameter, say for 
example $\Delta F_{ac}$ (or $C$, or whichever), 
and numerical integration during several periods ($T_b$) 
of the external driver. The convergence to the breather attractor 
corresponding to $F_{ac} + \Delta F_{ac}$ 
is guaranteed to be exponentially fast provided 
$\Delta F_{ac}$ is small enough and no bifurcation occurs.

The range of coupling values where these mobile solutions are 
observed is very far from the continuum regime; they are roughly 
in the range from $C=0.5$ to $1$, and its motion has a well defined 
average velocity. 
We should remark here that these MB's are not related to the 
breathers of the integrable (continuous) sine-Gordon model. 
The continuum limit of Eq. (\ref{eq:mot}) is the continuous 
sine-Gordon equation under external ac forcing and losses, which 
does not support mobile breather solutions \cite{quintero98}. 

A convenient quantitative descriptor of a mobile localized solution 
is its translation velocity, $v_b$. In order to define 
precisely this quantity, one has to introduce a continuous 
{\em collective variable} $X(t)$, naturally interpreted as 
the instantaneous center of localization of energy:

\begin{equation}
X = \frac {\sum_{j=-\infty}^{\infty} j\cdot
\tilde{e}_j}{\sum_{j=-\infty}^{\infty} \tilde{e}_j}
\label{eq:CM}
\end{equation}
where $\tilde{e}_j = e_j - e_{\infty}$ is the energy density 
referred to the background, {\em i.e.}: 

\begin{eqnarray}
e_j &=& \frac {1}{2} {\dot{u}_j}^2 +  \frac {1}{(2\pi )^2}
\big[1-\cos(2\pi u_j)\big] 
\nonumber
\\ 
&&+ \frac {C}{4}(u_j-u_{j-1})^2 +
\frac {C}{4}(u_{j+1}-u_{j})^2
\label{eq:ejota}
\end{eqnarray}
and $e_{\infty}$ is $e_j$ at a site $j$ far away from the 
exponentially localized breather core. The breather velocity 
is then defined as the following long-term average velocity:

\begin{equation}
v_b = \langle \dot{X} \rangle = \lim_{T\rightarrow\infty} 
\frac{1}{T} \int_{t_0}^{t_0 + T} \dot{X}\, dt
\label{velocity}
\end{equation}

Along a single branch of continued MB's, the breather velocity 
defines a continuous {\em but not necessarily smooth} curve. Let us 
emphasize that moving solutions of velocity $v_b$ posses two 
characteristic time scales, namely $\omega_b$ and $2 \pi v_b$. 
The issue of time scales competition is thus of concern here, 
in the sense that one would naturally expect the emergence of 
typically associated phenomena, like resonances and 
synchronization in the behavior of MB's, as demonstrated 
by numerical results to be shown later in section III. Now let us 
introduce some basic notions needed for what follows.

\subsection{Resonant mobile discrete breathers.}
\label{subsec:resonant}

A close view on Eq.(\ref{eq:mot}) reveals the basic 
symmetries of the dynamics: 

\begin{itemize}
    \item[(a)] Lattice translation (homogeneous system),
      \begin{equation}
      \mathcal{L}\{u_j(t)\} = \{u_{j+1}(t)\} 
      \end{equation}
    \item[(b)] Discrete time $T_b$ shift (periodic driver),
      \begin{equation}
      \mathcal{T}\{u_j(t)\} = \{u_j(t+T_b)\}
      \end{equation}
    \item[(c)] Space-time mirror symmetry (specific to such situations).
      \begin{equation}
      \mathcal{S}\{u_j(t)\} = \{-u_j\left(t+\frac{1}{2}T_b\right)\}
      \end{equation}
      This operator combine the mirror symmetry $\mathcal{R}(u) = -u$ of the
local term $\sin(u)$ and the center symmetry of the sinusoidal driver:
$\mathcal{S} = \mathcal{R} {\mathcal{T}}^{1/2}$
\end{itemize}

Given any solution $\{\hat{u}_n(t)\}$, the combined action of the symmetry
operators ${\mathcal{L}}$, ${\mathcal{T}}$ and ${\mathcal{S}}$ gives a
family of solutions generated by the symmetry group. We will see that the
existence and structure of this family plays a central role in our
interpretation of numerical results on desynchronization of moving DB's. 

Now we will make precise the notion of resonant DB, used in the introductory 
section. A ($p/q$)-resonant state is a synchronized orbit defined as fixed
point of the symmetry element ${\mathcal{L}}^{p}{\mathcal{T}}^{q}$, {\em i.e.}

\begin{equation}
\hat{u}_{n+p}(t+qT_b)=\hat{u}_n(t)
\end{equation}
If $p$ and $q$ have no common divisors, the state is said {\em pure
resonant}.  Note that it follows from the definition that a
($p/q$)-resonant MB has a velocity $v_b = (\omega_b/2\pi)(p/q)$. But
it is also important to realize that among all conceivable evolutions
$\{u_n(t)\}$ with this velocity, a resonant one is certainly very
special, for it posses a specific
(${\mathcal{L}}^{p}{\mathcal{T}}^{q}$) symmetry.

\subsection{Linear stability analysis.}
\label{subsec:Floquet}

Let us consider a small perturbation 
$\{\epsilon_j(t),\dot{\epsilon}_j(t)\}$ of a given  
DB solution $\{u_j(t)\}$. Linearizing around this 
solution, we obtain:
\begin{equation}
  \ddot{\epsilon}_j + \alpha\dot{\epsilon}_j +
  \cos(2\pi u_j(t))\epsilon_j = C \left(
    \epsilon_{j+1}-2\epsilon_j+\epsilon_{j-1} \right)
  \label{eq:linmot}
\end{equation}
An immobile DB of frequency $\omega_b$ is a fixed point of the operator 
${\mathcal{T}}$. The {\it Floquet} matrix maps a basis of the tangent 
space ($\{\epsilon_j(0), \dot{\epsilon}_j(0)\}$),
onto $\{\epsilon_j(T_b), \dot{\epsilon}_j(T_b)\}$, and it is given by 
the Jacobian of ${\mathcal{T}}$, {\em i.e.} $D({\mathcal{T}})$. The 
spectrum of eigenvalues of this Floquet matrix gives the linear stability 
of immobile DB's, allowing the characterization of the bifurcations that 
these solutions experience along continuation paths, as shown in Refs. 
\cite{martinez03,MFMF01}. In order to use these powerful Floquet 
methods for the analysis of mobile solutions, they must be periodic and so, 
one has to restrict attention to ($p/q$)-resonant MB's. 

Because a ($p/q$)-resonant MB is a fixed point of the operator 
${\mathcal{L}}^{p}{\mathcal{T}}^{q}$, the (extended) Floquet 
matrix providing the linear stability of a ($p/q$)-resonant MB is 
$D({\mathcal{L}}^{p}{\mathcal{T}}^{q})={\mathcal{P}}{\mathcal{M}}$, 
where ${\mathcal{M}}$ is the matrix of the linearized equations of motion
integrated over $q$ $T_b$-periods, and (periodic boundary conditions) 
${\mathcal{P}}$ is a cyclic permutation matrix of $p$ sites.
\begin{eqnarray}
\{\hat{u}_{j}(t_0)+\epsilon_{j}(t_0)
&,&
\hat{\dot{u}}_{j}(t_0)+\dot{\epsilon}_{j}(t_0)\} 
\\
&\rightarrow&
\{\hat{u}_{j}(t_0), \hat{\dot{u}}_{j}(t_0)\} 
+
{\mathcal{P}}{\mathcal{M}}\{\epsilon_j(t_0),\dot{\epsilon}_{j}(t_0)\}
\nonumber
\label{eq:monodromy}
\end{eqnarray}

The distinctive property of being an attracting solution of the nonlinear
evolution equation (\ref{eq:mot}) translates into the mathematical assertion
that an attracting $(p/q)$-resonant state has an associated Jacobian matrix 
$D({\mathcal{L}}^{p}{\mathcal{T}}^{q})$ with (bounded) spectrum inside the 
complex unit circle:
\begin{equation}
\sup |\mu| \leq 1
\end{equation}
where $\mu$ denotes eigenvalues of the Floquet matrix.

Eventually, the exit of a Floquet eigenvalue from the unit circle signals the
destabilization of the ($p/q$)-resonant MB by perturbations along the
associated Floquet eigenvector. In the linear regime these
destabilitations will grow with
exponential rate.

\section{The internal structure of the locking regime}
\label{sec:results1}

First of all, we briefly review the method used to generate mobile
discrete breathers (MB). We begin by generating an immobile discrete 
breather, starting from the anti-continuum limit $(C=0)$. Initially, 
we have used the same parameters as in \cite{MFMF01}: $F_{ac}=0.02$, 
$\omega_{b} =0.1\times2\pi$ and $\alpha=0.02$. As we increase $C$ 
adiabatically, the discrete breather solution remains as an attractor 
of the dynamics \cite{MFMF01}, enabling us to find breathers at 
different values of $C$, by continuation.

We continue the breathers until  the first pitchfork bifurcation
\cite{CAT96,AC98,MFMF01}, which connects one-site breathers 
with two-site breathers via asymmetric ones.
The localized eigenmode, responsible for this instability is
asymmetric, and can be used to ``depin'' the static breather. 
Then, we generate a MB by perturbing the static breathers
along this mode with an amplitude $\mu$. As in
\cite{CAT96,MFMF01}, we found MB if the perturbation is larger
than some critical value ${\mu}_c$. Unlike Hamiltonian systems, 
the velocity reached by the MB is independent of $\mu$  for 
$\mu > {\mu}_c$. This method allows us to produce MB in 
a wide range of the coupling parameter $C$.
Two kinds of MB's have  been observed depending on $C$:
for $C$ in the interval [0.5,0.89] only induced MB's exist in
coexistence with static breathers. However, in the $C$ range
[0.89,0.97] spontaneous MB's appear as attractor solutions, coexisting
whith the induced MB's. Static breathers are not found in this range.
Hence, we have decided to focus our research on some points
in those regions. The selected values are $C=0.55$ and $C=0.75$,
in the first region and $C=0.94$ in the second.

In order to check the dependence of these MB's on the driving force,
we fix $C$ and then we vary $F_{ac}$. Some general features emerge
in all cases. The simulations show the appearance of steps with
velocities $v_b = (\omega_b/2\pi)(p/q)$. Recall that these
velocities are the velocities of mode-locked MB. But this does 
not guarantee that the MB inside the locking step is 
($p/q$)-resonant, so we must check the periodicity of these MB's 
inside these steps. Another empirical observation is that
the limit value of $F_{ac}$ before the destruction of the MB,
increases with $C$.

\begin{figure}
  \begin{center}
    \includegraphics[width=0.4\textwidth]{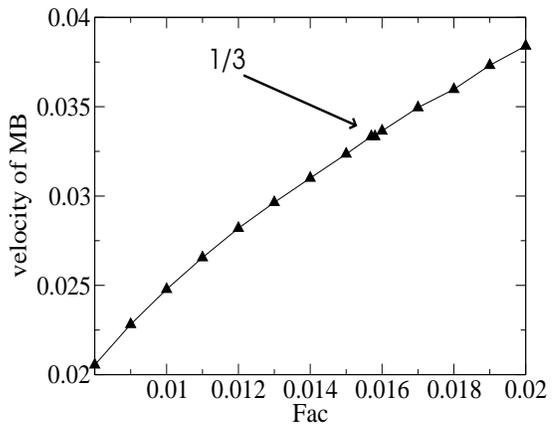}
    \caption{Velocity of the MB vs. $F_{ac}$ for $C=0.55$.
    A very narrow plateau $\frac{1}{3}$ appears. The rest of fixed
    parameters are $\omega_{b} =0.1\times2\pi$ and $\alpha=0.02$}
    \label{fig1}
  \end{center}
\end{figure}

The results with $C=0.55$  are sketched in fig. \ref{fig1}. 
This figure shows the full $F_{ac}$ range in which MB's with a 
definite velocity exist. One can see an extremely narrow step at 
velocity $v_b = (\omega_b/2\pi)(1/3)$. The Poincar{\'e} section of 
these solutions reveals that they are {\it pure resonant MB's}. Moreover, 
the linear stability analysis reveals that they are linearly stable 
inside the whole step interval, i.e. $[0.01568, 0.01581]_{F_{ac}}$. 
This type of solution, i.e. ($1/3$)-resonant, has been found in the 
whole range of $C$ values that we investigated.

\begin{figure}
  \begin{center}
    \includegraphics[width=0.4\textwidth]{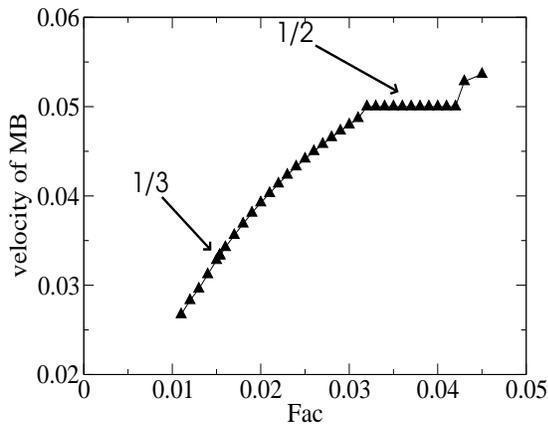}
    \caption{Velocity of the MB vs. $F_{ac}$ showing two mode-locking
    plateaux ($\frac{1}{3}$ and $\frac{1}{2}$) for $C=0.75$, the
    values of $\gamma$ and $\omega_b$ are the same that in fig. \ref{fig1}}
    \label{fig2}
  \end{center}
\end{figure}

For $C=0.75$, the results of the velocity-force curve are represented
in the Fig. \ref{fig2}. In this case we find a bigger step at velocity
$v_b = (\omega_b/2\pi)(1/2)$ ($[0.03177,0.042]_{F_{ac}}$), and, again,
a very little one at $v_b = (\omega_b/2\pi)(1/3)$
($[0.01534,0.01538]_{F_{ac}}$).  In the whole interval of this
last step, the MB is ($1/3$)-resonant and linearly stable.  On the
other hand, the bigger step has a more complex structure: In the range
$[0.03177,0.03831]_{F_{ac}}$, the MB is periodic with period
$2T_{b}$ and linearly stable.  These ($1/2$)-resonant solutions can be
continued in coupling parameter from $C\simeq 0.71$ up to $C\simeq
0.84$.  However at $F_{ac}\simeq 0.03831$ the MB suffers a transition
of period tripling (a Floquet eigenvalue and its complex conjugate
cross the unit circle at angles $2\pi/3$ and $-2\pi/3$), and rapidly
goes to a chaotic state via a {\em period doubling cascade}, for
$F_{ac}\simeq 0.03833$. The MB remains in this chaotic state with
commensurate velocity, and around $F_{ac}\simeq 0.042$, the solution
leaves the step. 

\begin{figure}
  \begin{center}
    \includegraphics[width=0.4\textwidth]{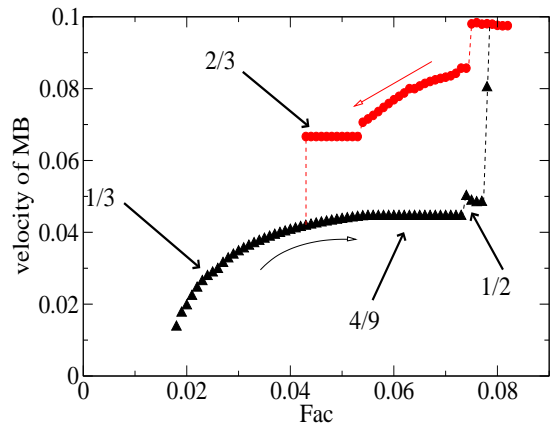}
    \caption{Same as figures \ref{fig1} and \ref{fig2} now for $C=0.94$. 
      Several mode-locking steps are visible. Note, also
    the hysteresis in the curve.}
    \label{fig3}
  \end{center}
\end{figure}

Finally, at $C=0.94$ we obtain the curves represented in the Fig.
\ref{fig3}. There we can observe a very rich behavior of the breather
velocity: different steps at velocity values of
$(\omega_b/2\pi)(1/3)$,$(\omega_b/2\pi)(4/9)$ and
$(\omega_b/2\pi)(2/3)$, as well as an evident hysteresis. The latter
is somehow typical of underdamped systems; it implies the coexistence
of (at least) two different MB attractors, with the same model
parameters which has been reported previously
\cite{MFMF01,MATTHIAS}. The dynamics in these steps is simple:
periodic and linearly stable, with no bifurcations inside the
plateau. Also a smaller plateau with velocity $(\omega_b/2\pi)(1/2)$
is found, although in this case only quasiperiodic MB's exist.  For
the sake of completeness we mention that the upper branch can be
continued to lower values of the coupling and connects with MB's
generated by the depinning mode, whereas the lower branch belongs to
the spontaneous MB.

Summarizing, the step with velocity
$(\omega_b/2\pi)(1/3)$ is found in all the range of coupling where
MB's seem to exist. The range for the step solutions with 
$v_b =(\omega_b/2\pi)(1/2)$ starts at $C=0.71$ since 
below this coupling a MB with regular motion does not
exist, in the parameter range studied. At higher $C$, 
we observe other locking steps like $v_b = (\omega_b/2\pi)(4/9)$
and $v_b =(\omega_b/2\pi)(2/3)$ for $C=0.94$. Although we have not
been exhaustive varying in the parameter $C$ we can certainly find these
steps (and others) in the whole range of coupling parameters, where DB
are found.

Some of the observed locking plateaus coincide with the stability 
interval of the corresponding ($p/q$)-resonant MB, but in other 
cases (notably the $1/2$ locking step shown in Fig. \ref{fig2}) 
the resonant state destabilizes inside the plateau and the new 
attracting solution (with locking velocity $v_b$) is more complex: 
either periodic with a larger period, or even chaotic, as revealed 
by the computed (largest) Lyapunov exponent. These types of complex 
behaviors inside a locking plateau are known to happen for a single 
driven-damped anharmonic oscillator \cite{baker}, as well as for moving 
discommensurations \cite{MARTINEZ97}. In this regard, the result for 
discommensurations can be reducible to the single-particle case using 
collective variable approaches. In our case, the breather internal degrees
of freedom, together with the $X$ (breather center) variable, do not allow
for a straightforward reduction to single-particle behavior.  


\section{Unlocking transition}

A typical route for the transition from a periodic state to a
quasi-periodic or chaotic one in dissipative systems is that 
mediated by intermittencies \cite{BERGE84}. Intermittencies occur 
whenever the behavior of a system seems to switch  between 
qualitatively different unstable periodic orbits or behaviors 
(periodic, quasi-periodic or chaotic) even though all the control 
parameters remain constant and no external noise is 
present \cite{HILBORN94}. Depending on the type of Floquet 
instability of the periodic orbit responsible for the bifurcation 
(crossing the unit circle at $+1$, at two complex conjugate 
eigenvalues or at $-1$) intermittencies are classified into 
intermittencies type I, II, or III, respectively \cite{BERGE84}.

In this section we study the unlocking transition of the 
($p/q$)-resonant MB and characterize it as a transition from 
periodicity to quasi-periodicity driven by the regular (periodic) 
appearance of type I intermittencies. This mechanism for unlocking 
transitions was already observed in the purely dissipative 
dynamics of ac driven modulated structures of the FK model 
\cite{FALO93}. Intermittencies of type I are also known to be 
responsible for the depinning transition of discrete solitons 
(discommensurations) of the underdamped FK model \cite{MARTINEZ97}.

From now on in this section we will focus our attention on the unlocking 
transition at the left edge of the $1/2$ locking step of Fig. \ref{fig2} 
($C = 0.75$) which occurs at $F_{ac} = 0.03177$. The extended Floquet
analysis of the ($1/2$)-resonant MB close to the edge reveals that an 
eigenvalue of the Floquet matrix $D({\mathcal{L}}{\mathcal{T}}^{2})$ 
approaches the value $+1$ from the interior of the complex unit circle 
along the real axis, and leaves the unit circle at the transition point. 
The eigenvector associated to this Floquet eigenvalue is exponentially
localized at the breather center. Out the step, the ($1/2$)-resonant 
MB is thus linearly unstable.

\begin{figure}
  \begin{center}
    \includegraphics[width=0.4\textwidth]{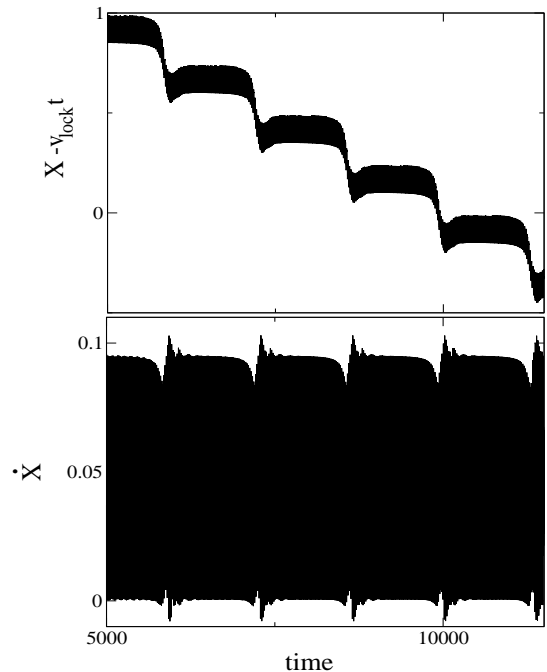}
    \caption{Upper panel: $X - v_{lock}t$,  i.e. $X$ respect to a
      moving frame that moves with $v_b = (\omega _b/2\pi)1/2$ just
below the left edge of the $\frac{1}{2}$ step of the Fig.\ref{fig2}.
Jumps correspond to the intermittencies described in the text. Lower panel:
$\dot{X}$ in the same point of response curve.}
    \label{fig4}
  \end{center}
\end{figure}

In order to visualize the effect of this instability we plot in Fig.
\ref{fig4} the breather center (out of the step but very close to its
edge) $X$ and its velocity $\dot X$ in a reference frame moving with
the locking velocity $(\omega_b/2\pi)(1/2)$. One can clearly see in
figure \ref{fig4} that the breather center remains for some time
intervals in laminar regimes (of locked velocity), interrupted by
sudden jumps of very short duration.

We have numerically checked that these laminar regimes correspond to
the discrete family of equivalent unstable continuations of the
($1/2$)-resonant MB, related one to each other by symmetry operations
$\mathcal{L}$, $\mathcal {S}$ and $\mathcal {T}$.  Therefore, the
destabilizing Floquet eigenvector pushes the weakly unstable resonant
MB $\hat{u}$ towards its (equivalent) closest member of the family
(that turns out to be $\mathcal {S}^{-1}\hat{u}$), which is also
unstable, and so on. The duration of each laminar regime, which
diverges at the bifurcation point, is proportional to $(\mu
-1)^{-1}$, where $\mu$ is the unstable Floquet eigenvalue.

\begin{figure}
  \begin{center}
    \includegraphics[width=0.4\textwidth]{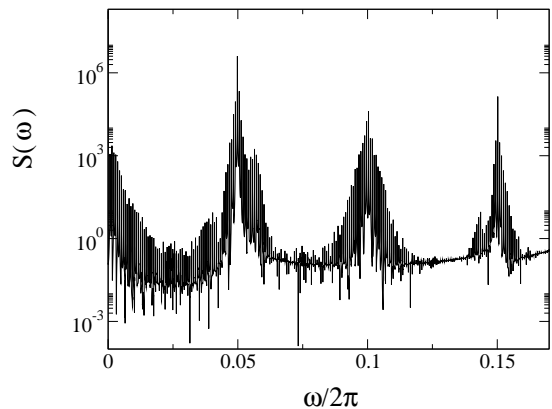}
    \caption{ Power spectrum $S(\omega)$ of $\dot{X}$ showing
clearly a quasiperiodic behavior.}
    \label{fig5}
  \end{center}
\end{figure}

The computation of the {\it Power Spectrum} of $ \dot X $, i.e,
\begin{equation}
\label{power_spectra}
S(\omega)
=
\left |
\int_{-\infty}^{\infty}
\dot X
(t)
e^{i\omega t}
dt
\right |
^2
\end{equation}
for the attracting MB out of the locking plateau (see figure \ref{fig5}), 
reveals the new frequency $ \omega_{int} = \mu -1$ associated to the 
intermittencies, and further confirms that they appear at regular time
intervals, so that the attracting MB out of the step is quasiperiodic. 

This scenario of unlocking mechanism is confirmed to happen for other
numerically obtained plateaus of mode-locked velocity. It appears that 
the desynchronization by regular intermittent phase shifts pulses is quite a
generic phenomenon.

\section{The effects of subharmonic perturbations of the driving force}

In the previous sections, we have shown how complex the dynamical
response of the system can be. This includes, among others, resonant
MB's which are destabilized via intermittencies when the parameters are
changed.  Inside the resonant step, the particular behavior of the
oscillators forming the breather can be complex (highorder periodic
or chaotic). However it does not prevent the breather to have a
definite mean velocity, commensurate with the external frequency.

Our goal in this section is to enlarge the regions of the parameter
space for which those resonant steps exist. The structure of the
quasiperiodic attractor in the vicinity of the resonant steps gave us
some indications about the procedure to follow. We must reinforce the
laminar phase (correspondingly inhibit the appearance of
intermittencies) applying a perturbation subharmonically related with
the original ac force:
\begin{equation}
  F(t) = F_{ac}\sin(\omega_b t) + 
\lambda \sin \big(\frac{\omega_b}{n}t + \Delta \big)
  \label{eq:second}
\end{equation}
where $\lambda$ is small compared to $F_{ac}$, $n$ is a positive
integer and $\Delta$ represents a phase shift between
both terms.

\begin{figure}
  \begin{center}
    \includegraphics[width=0.4\textwidth]{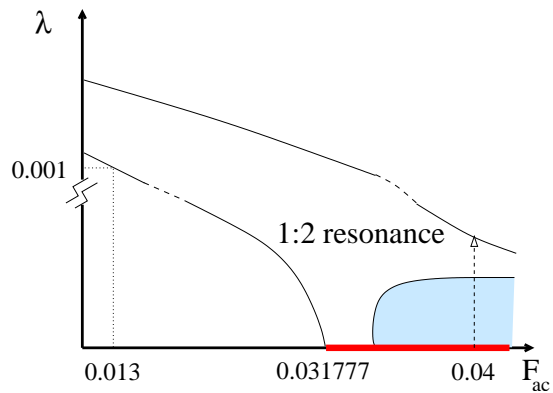}
    \caption{Schematic phase diagram of the behavior with a
subharmonic perturbation of strength $\lambda$ vs. unperturbed driving
force amplitude $F_{ac}$. Shaded region refers to more complex behavior (chaotic,
quasiperiodic or higher resonances than $1/2$) but also mode-locking.
    The arrow stands for the path followed in Fig. \ref{fig7}}
    \label{fig6}
  \end{center}
\end{figure}

Such kind of perturbation has been proved to be efficient to stabilize
linearly unstable periodic orbits. It has been used in systems with
few degrees of freedom \cite {braiman, barbi-salerno} as well as in
solitons \cite {salerno} and even in experiments \cite
{VFB95}. In our case we achieve to lower the onset of the resonant
steps significantly. In particular, we focus our attention on the step
$1/2$ in $C=0.75$ which presents the richest phenomenology, and choose
$n=2$ in (\ref{eq:second}) and $\Delta=0$ (other $\Delta$ values have
been used and the results obtained are essentially similar). The main
results are summarized in the phase diagram of Fig. \ref{fig6}. For
example the onset of the resonant step is reduced from
$F_{ac}=0.031777$ to $F_{ac}\simeq 0.013$ when $\lambda$ increases
from $0$ to $\lambda \simeq 0.001$.

 This effect adds to another that takes place inside the step and that
 is related with the control of chaos. A chaotic attractor
 like the one developed by a period doubling cascade inside the 1/2
 resonant step for values $F_{ac}\approx 0.03833$ (with $\lambda=0$) is
 formed by a dense set of unstable periodic orbits of different
 periodicities but all of the same velocity. The addition of a
 suitable perturbation is able to stabilize one of these unstable
 periodic orbits.  The
 taming of chaotic states in nonantonomous dynamical systems by the
 addition of harmonic perturbations is a quite general phenomenon
 \cite{braiman, salerno,Chacon02}, and we observe it in the case of these
 chaotic breathers with locked velocity.

\begin{figure}
  \begin{center}
    \includegraphics[width=0.4\textwidth]{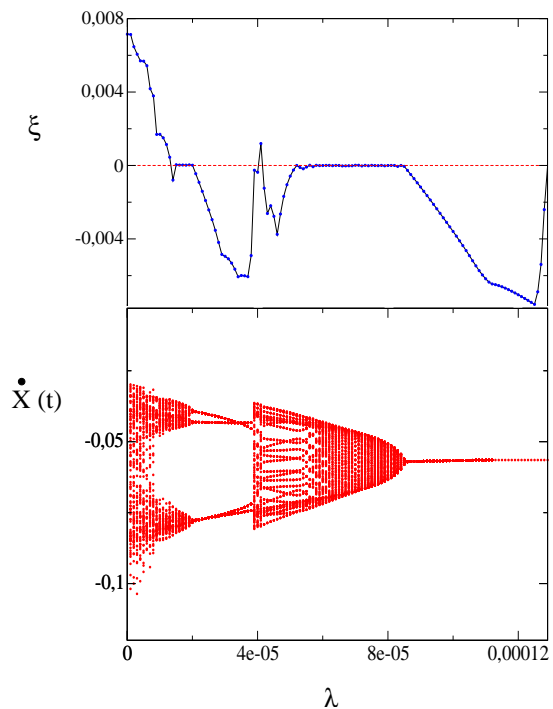}
    \caption{(upper) Largest Lyapunov exponent ($\xi$) for $F_{ac}=0.04$. 
             (lower) Poincar\'e section of $X$ at intervals $2T_b$, so a 
	     single point is indicative of a periodic solution 
in the mode-locking step at its corresponding $\lambda$ value. }
    \label{fig7}
  \end{center}
\end{figure}

To quantify this behavior we compute the largest Lyapunov exponent
\cite{method} of the MB solution at a fixed $F_{ac}= 0.04 $ as a
function as the perturbation strength $\lambda$. We start at a chaotic
mode-locking MB. As soon as we apply the perturbation, a significant
decrease of the Largest Lyapunov exponent is observed until a narrow region of
quasiperiodicity is reached (Largest Lyapunov exponent $=0$). Then, a
sequence of periodic and quasiperiodic solutions follows and finally,
a broad region of periodic solution, with the
perturbation period (see Fig. \ref{fig7}).  Note that all this is
attained with a  $\lambda$  two orders of magnitude
smaller than $F_{ac}$.

At fixed $C$, one can \emph{tune}, by varying $\lambda$, the desired
periodic state. This method is extremely robust against changes in
$C$. The pure resonant MB solution can be extended for a wide range of the
coupling $C$. Using an appropriated $\lambda$ we have extended the
$1/2$-resonant solution from $[0.71, 0.84]$ to $[0.5, 0.84]$ in the
coupling parameter.

\section{Concluding remarks and Summary}

We have studied mobile discrete breathers in the underdamped
Frenkel-Kontorova model.  Periodic (or $p/q$-resonant) solutions have
been found for a large range of parameter values (as $C$ and
$F_{ac}$).  These states are structurally stable as they are a
consequence of the synchronization between the two time scales of the
mobile breather. This synchronization (or its absence) results in a very rich
dynamics of the mobile breather solutions, including quasiperiodic and chaotic
ones. Localized chaotic behavior has been previously observed in
static breathers \cite{Martinez99} but, here we show that it is
compatible with mode-locking motion of the breather center (core).

One important issue in the study of localized discrete excitations is
the possibility of their dynamical description by reduction to a
system with few degrees of freedom (or collective coordinates). Up to
our knowledge, two collective coordinate scheme have been developed
with mobile breathers.  The first approach described in \cite{willis}
has got very fruitful results for discrete \cite{MARTINEZ97}
and continuum  \cite{morales} solitons, as well as for continuum breathers
\cite{sanchezbishop,caputo}.  This approach uses as starting point for
the calculations the continuum model solution and fails far from this
limit.

The second approach \cite{mackay02} is the geometric counterpart of
the first.  
It works with mobile breathers in a hamiltonian framework, which
prevents its direct use in the dissipative case considered here.
However, the possibility of its use in non-hamiltonian contexts, under
certain technical conditions, is still open as was briefly pointed out
in ref. \cite{mackay03}

Although analytical methods for a collective coordinate approach to
mobile dissipative discrete breathers have not been yet developed, our
numerical results strongly suggest that such an approach could give an
accurate account of most of the observed phenomenology. We hope that
the results presented here could encourage for the search of those
methods.

Finally, we remark the important role that a second harmonic in the
forcing plays in the dynamics of the MB.  We observe that the presence
of a small subharmonic driving term enhances $(p/q)$-resonant
solutions.  Increasing the subharmonic amplitude (\ref{eq:second})
from zero, an initially chaotic MB can be tamed, above some threshold,
obtaining a non-chaotic one (quasiperiodic or $(p/q)$-resonant). The
feasibility of the experimental implementation of different wave forms
for driving forces could falicitate the observation of these solutions
since they are more robust and stable.

In summary, we have shown that mode-locking motion of breathers are
stable solutions for an important example of non-linear lattice such as the
Frenkel-Kontorova model. We have characterized this mode-locking solution
as well as the mechanisms for the unlocking. Finally we have applied a simple
method to extend this kind of solutions to a large range of values
in parameter space.

\section{Acknowledgements}

We thank J. L. Garc\'{\i}a Palacios and J. J. Mazo for the reading of the
manuscript.  DZ is grateful to M. Meister for discussions.  Financial
support is from European Network LOCNET HPRN-CT-1999-00163 and also
from Spanish MCyT and European Regional Development Fund (FEDER)
program through BFM2002-00113 project are acknowledged.  DZ also
acknolewdges a DGA fellowship.


\ifx\undefined\allcaps\def\allcaps#1{#1}\fi\ifx\undefined\allcaps\def\allcaps#1{#1}\fi

\end{document}